\documentclass[aps,
preprint,
superscriptaddress,groupedaddress]{revtex4} 
\usepackage{latexsym}       
 \usepackage{natbib}
\usepackage{graphicx}  
\usepackage{anysize} 
\usepackage{array}
\usepackage{float}
\usepackage[T1]{fontenc}
  
\begin{document}

\title{Interacting social processes on interconnected networks}

\author{Lucila G. Alvarez Zuzek} \email{lgalvere@mdp.edu.ar}
\affiliation{IFIMAR, Instituto de
  Investigaciones F\'{i}sicas de Mar del Plata (CONICET-UNMdP), 7600
  Mar del Plata, Argentina}

\author{Cristian E. La Rocca} \affiliation{IFIMAR, Instituto de
  Investigaciones F\'{i}sicas de Mar del Plata (CONICET-UNMdP), 7600
  Mar del Plata, Argentina}

\author{Federico Vazquez} \affiliation{IFLYSIB, Instituto de
  F\'{\i}sica de L\'{\i}quidos y Sistemas Biol\'ogicos (CONICET-UNLP),
  1900 La Plata, Argentina}

\author{Lidia A. Braunstein} \affiliation{IFIMAR, Instituto de
  Investigaciones F\'{i}sicas de Mar del Plata (CONICET-UNMdP), 7600
  Mar del Plata, Argentina}

\begin{abstract}

We propose and study a model for the interplay between two different
dynamical processes --one for opinion formation and the other for
decision making-- on two interconnected networks $A$ and $B$.  The
opinion dynamics on network $A$ corresponds to that of the M-model,
where the state of each agent can take one of four possible values
($S=-2,-1,1,2$), describing its level of agreement on a given issue.
The likelihood to become an extremist ($S=\pm 2$) or a moderate
($S=\pm 1$) is controlled by a reinforcement parameter $r \ge 0$.  The
decision making dynamics on network $B$ is akin to that of the
Abrams-Strogatz model, where agents can be either in favor ($S=+1$) or
against ($S=-1$) the issue. The probability that an agent changes its
state is proportional to the fraction of neighbors that hold the
opposite state raised to a power $\beta$.  Starting from a polarized
case scenario in which all agents of network $A$ hold positive
orientations while all agents of network $B$ have a negative
orientation, we explore the conditions under which one of the dynamics
prevails over the other, imposing its initial orientation. We find
that, for a given value of $\beta$, the two-network system reaches a
consensus in the positive state (initial state of network $A$) when
the reinforcement overcomes a crossover value $r^*(\beta)$, while a
negative consensus happens for $r<r^*(\beta)$. In the $r-\beta$ phase
space, the system displays a transition at a critical threshold
$\beta_c$, from a coexistence of both orientations for $\beta<\beta_c$
to a dominance of one orientation for $\beta>\beta_c$. We develop an
analytical mean-field approach that gives an insight into these
regimes and shows that both dynamics are equivalent along the
crossover line $(r^*,\beta^*)$.
\end{abstract}

\maketitle

\section{Introduction}
\label{intro}

The study of complex networks has become a matter of great interest to
scientists, due to the large number of real systems that evolve on top
of these kind of topological structures, such as human societies,
climate, transportation and physiological systems.  For many years
researchers were focused on studying the topology of isolated
networks, and its effect on different dynamics
\cite{New_10,Cohen_10,Bara_02,Pastor_15,Vazquez_08,Vazquez_10,Sanmiguel_09,Lar_09,Ana_01,Val_11,Bashan_01,Goz_08,Dan_14}.
However, it is known that many real-world systems are not isolated but
they interact with each other, and they are well described by a
multilayer system of interconnected networks
\cite{Domenico_13,Boccaletti_14,Jianxi_10,Perc_15}, where nodes
belonging to different networks interact.  A different multilayer
context is that of multiplex networks, in which the same nodes exist
--and represent the same entity-- in different network layers (see
\cite{Boccaletti_14} and references therein).  The study of multilayer
systems allows to understand the interplay between complex networks,
and how this affects the processes propagating on them, e.g,
synchronization \cite{Gambuzza_15,Torres_15}, diffusion
\cite{Gomez_13}, percolation
\cite{Bul_01,Gao_12,Val_13,Hackett_16,Baxter_16,Dimuro_16} and
epidemic spreading
\cite{Buono_14, Buono_15,AlvarezZuzek_15,Cozzo_13,Granell_13_1,Alvarez-Zuzek_15,Saumell_12,Vazquez_16,Scoglio_14}.
Within the context of social science, the study of social phenomena on
multilayers is relatively new \cite{Boccaletti_14}.  Multilayer
networks have recently been applied to study opinion dynamics
\cite{Fort_01}, a topic that has many analogies with the dynamics of
species competition \cite{Galam98}, and that has been extensively
studied by statistical physicists.  In reference \cite{Halu_13}, Halu
et al. use two interacting networks to describe two political parties
that compete for votes in an election.  Diakonova et al. explored in
\cite{Diakonova_14} the dynamics of the voter model for opinion
formation on a bilayer network system with coevolving links, and also
studied in \cite{Diakonova_16} the reducibility of the voter model on
a two-layer multiplex to a single layer system.

The process of opinion formation may affect and depend on other
social processes like decision making \cite{Galam97}, due to the
relationships between the individuals taking part in each of these two
processes.  For instance, people in a civil society discuss and form
their opinions on a given issue, such as the legalization of the
marriage between people of the same sex. However, the decision on
whether the same-sex marriage law is approved or not is discussed and
finally taken in a legislative body, such as the Congress.  As a
consequence, these two social groups --society and Congress--
influence each other, as congressmen form part and interact with
members of the society and, at the same time, people in the society
are influenced by what the Congress is deciding.

In this article we investigate the interaction between two social
dynamics, one for opinion formation and the other for decision making,
that take place on two interconnected networks. The dynamics for
opinion formation corresponds to that of the model proposed by La
Rocca {\it et. al}~\cite{Larocca14}, to which we refer as the
M-model. This model possesses $2M$ different states describing the
spectrum of possible opinion orientations on a given issue, from
totally against (state $S=-M$) to totally in favor ($S=M$), with some
moderate opinions between these extreme values.  The M-model explains
the phenomena of polarization in a population of individuals that
evolve under pairwise interactions, by implementing two main social
mechanisms for opinion formation, compromise and persuasion
\cite{Mas_13,Mas_13b,Balenzuela_15}. The decision making dynamics is
akin to that of the Abrams-Strogatz (AS) model \cite{AS_03,Vazquez_10}
(originally introduced to study language competition), where agents
can choose between only two possible choices, to be either in favor
($S=+1$) or against ($S=-1$) the issue. Each agent may change its
decision by a mechanism of social pressure, in which the probability
of switching its present choice increases non-linearly with the number
of neighbors that make the opposite choice. In this work, we set the
system to explore a hypothetical polarized scenario where, initially,
all the agents in the opinion network are in favor of the issue
(positive orientations), while all the agents in the decision network
are against (negative orientations). By means of this simple model we
address the following questions: under which conditions the opinion
dynamics is able to influence and reverse the initial orientation of
the decision network?  Which dynamics is stronger and prevails in the
long run? We need to mention that the present proposed model on two
interacting networks has some analogies with models of coupled spin
systems previously studied to describe the phase diagram of
orientational glasses \cite{Galam87,Galam89}. We also notice that,
even though we use in this study the M-model and the AS model for
their simplicity, other social models can be implemented as well to
explore the interplay between opinion and decision making processes.

The rest of the paper is organized as follows. In Section \ref{model}
we introduce the model, describing the topology of interactions as
well as the dynamics that runs over each network. Results from
numerical simulations of the model are presented in Section
\ref{simulation}, where we show that there are three possible final
states: a coexistence of both orientations (neither dynamics
dominates), a positive consensus (opinion dynamics domination) and a
negative consensus (decision dynamics domination). Then, in Section
\ref{MF} we develop a mean field approach that allows to explain the
qualitative behavior of the system, and shows that both dynamics
behave equivalently for some particular choice of the
parameters. Finally, in Section \ref{conclusion} we summarize and
discuss our findings.

\section{The Model}
\label{model}

In our model we consider two interconnected networks, denoted by
networks $A$ and $B$, each with the same number of nodes $N$ and
intranetwork degree distribution $P(k)$, which represents the fraction
of nodes connected to $k$ other nodes within the same network.  We
also consider pairwise interconnections, that is, each node is
connected to one randomly chosen node in the other network, through an
internetwork link. Therefore, a node with $k$ intranetwork links and
one internetwork link is connected to a total of $k+1$ neighbors: $k$
from the same network and $1$ from the other network. In order to keep
the internetwork topology as simple as possible, we allow each node to
have only one internetwork link. However, the qualitative behavior of
the system is expected to be the same if other more complex
internetwork patterns are used.  In this particular topology, nodes
and links represent agents and their social interactions,
respectively, and thus the terms "nodes" and "agents" are used
alternatively along the article.

\begin{figure}[H]
  \centering
  \includegraphics[width=0.75\textwidth]{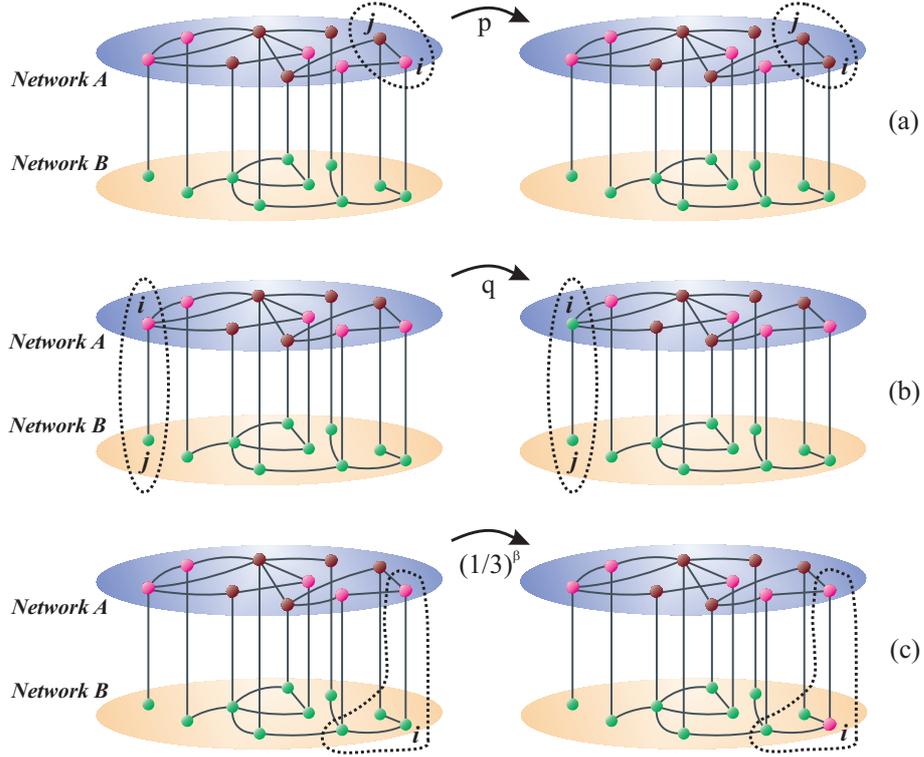}
  \vspace{0.4cm}
  \caption{\footnotesize
    Schematic representation of two interconnected networks
    with $N=10$ nodes in each layer. The dynamics on the top network
    $A$ (blue) obeys the M-model, while the dynamics on the bottom
    network $B$ (beige) is akin to that of the Abrams-Strogatz
    model. The colors of the nodes correspond to different opinion
    states: $S=1$ (pink), $S=2$ (burgundy) and $S=-1$ (green). The
    figures from the left (right) represent the situation before
    (after) the chosen node changes its state. (a) A moderate node $i$
    ($S_i=1$) and a extremist neighbor $j$ ($S_j=2$) in network $A$
    are chosen. Then $i$ becomes extremist with probability $p$
    ($S_i=1 \to 2$). (b) A moderate positive node $i$ ($S_i=1$) in
    network $A$ and a negative neighbor $j$ ($S_j=-1$) in network $B$
    are chosen. Then $i$ becomes a negative moderate with probability
    $q$ ($S_i=1 \to -1$). (c) The chosen node $i$ belongs to network
    $B$ and is a negative moderate ($S_i=-1$) with total degree
    $k_i=3$ (internal and external degrees $k=2$ and $k=1$,
    respectively). Then it changes orientation ($S_i=-1 \to 1$) with
    probability $(1/3)^{\beta}$.}
  \label{esquema}
\end{figure}

The dynamics on network $A$ corresponds to that of the M-model
\cite{Larocca14} with $M=2$, where only one random agent updates its
state at each time step, unlike the original version of the model
where two randomly chosen agents can change their states.  The opinion
state of each agent is represented by an integer number $S^A$ with
four possible values $S^A=-2,-1,1$ or $2$, where the sign of $S^A$
indicates its opinion orientation and its absolute value $|S^A|$
measures the intensity of its opinion.  Thus, $S^A=2$ and $S^A=-2$
represent positive and negative extremists, that is, people totally in
favor or against the issue, respectively, whereas $S^A=1$ and $S^A=-1$
describe moderate opinions from each side.  In a single step of the
dynamics, an agent and one of its neighbors are chosen at random.  A
moderate agent is persuaded by a same-orientation neighbor to become
an extremist with persuasion probability $p$ ($|S^A|=1 \to 2$
transition), while an extremist agent becomes moderate ($|S^A|=2 \to
1$) and a moderate agent changes orientation ($S^A=\pm 1 \to \mp 1$)
with compromise probability $q$ when they interact with an
opposite-orientation neighbor [see Figs \ref{esquema}A and
  \ref{esquema}B]. As we choose $p+q=1$ and the M-model dynamics
depends on the relative ratio $r \equiv p/q$ between the probabilities
to become an extremist or a moderate \cite{Larocca14}, we can express
both probabilities $p=r/(1+r)$ and $q=1/(1+r)$ as function of $r$. The
parameter $r$ measures the strength of \emph{reinforcement} in the
opinion orientation, i e., the tendency of same-orientation neighbors
to adopt a more extreme viewpoint as they persuade each other. Thus,
for large values of $r$ most agents tend to keep their opinions close
to the extreme values $S=2$ or $S=-2$, while for small $r$ opinions
tend to remain close to the moderate values $S=1$ or $S=-1$. This
model was studied on single fully connected networks in
\cite{Larocca14}, where it was shown that the system reaches a
quasistationary state whose features depend on $r$.  A polarized state
is obtained for $r>1$ (persuasion larger than compromise), where
agents' opinions are driven to the extreme values $M$ and $-M$, and
thus the distribution of opinions becomes "U-shaped", with peaks at
$M$ and $-M$. A centralized state is observed for $r<1$ (compromise
larger than persuasion), in which most agents hold opinions close to
the moderate values $1$ and $-1$. The final state in the long time
limit corresponds to an opinion consensus in either state $M$ or $-M$
(all agents in the same state $M$ or $-M$), depending on whether there
is an initial majority of positive or negative agents,
respectively. When the system reaches this completely ordered state
opinions cannot longer evolve, and thus we say that consensus is an
absorbing state of the dynamics.

The decision making dynamics of network $B$ is similar to that of the
AS model \cite{AS_03,Vazquez_10}, where each agent can choose to be
either in favor (choice state $S^B = +1$) or against (choice state
$S^B = -1$) the given issue. This non-linear version of the voter
model \cite{Vazquez_08} implements the peer pressure as a social
mechanism to change an attitude or behavior: an agent can change its
mind and reverse its decision with a probability equal to a power
$\beta$ (the \emph{volatility}) of the fraction of its opposite-choice
neighbors [see Fig \ref{esquema}C]. The volatility exponent
$\beta$ measures how prone a node is to changing state, from very
likely for $\beta \simeq 0 $ to very unlikely for $\beta \gg 1$.  The
dynamics of the AS model was extensively studied in single topologies,
including fully connected networks as well as complex networks and
lattices (see \cite{Vazquez_10} and references therein).
This model exhibits a transition from a coexistence of both states
(even mix of $+1$ and $-1$ agents) to a consensus in either state $+1$
or $-1$, as $\beta$ overcomes a threshold value $\beta_c \simeq 1$
that is slightly sensitive to the topology of interactions and the
symmetry between both states. The coexistence regime of non-consensus
is quasistationary in finite systems, because finite-size
fluctuations eventually drive the system to one of the two absorbing
consensus states.

A distinctive feature of both the M-model and the AS model on single
topologies is that their consensus states are attractive.  Therefore,
starting from a configuration where all agents have the same state
$S=\pm M$ in the M-model (or $S=\pm 1$ in the AS model), we can
introduce a small perturbation by changing the states of a few agents
at random, and check that the dynamics quickly brings the system back
to the initial consensus state.  The stability of the consensus state
in the M-model increases with $r$, as agents have a larger probability
to adopt and keep their initial extreme opinions.  For its part, the
stability of consensus in the AS model increases with $\beta$, as
agents are less likely to change their choices.  Then, an interesting
situation happens when these two models are coupled and start from
opposite oriented consensus states, given that each dynamics tries to
bring the entire two-network system to its own initial state.  The
interplay between the two dynamics would eventually drive the system
to one of the two initial consensus states, and thus we can interpret
this outcome as the prevalence of one dynamics over the other.  We
expect that the final result depends on the relative values of
parameters $r$ and $\beta$, which are proportional to the "strength"
of the M-model and the AS model, respectively.

Since we are interested in studying which dynamics dominates in the
long run we initially set all nodes in network $B$ to state $S^B=-1$,
while in network $A$ we randomly assigned state $S^A=2$ to $N/2$ nodes
and state $S^A=1$ to the other $N/2$ nodes (all nodes positively
oriented but with different intensities). Then, at each time step of
length $\Delta t=1/2N$, a node $i$ is chosen at random from the two
networks and its state $S_i$ is updated according to whether $i$
belongs to network $A$ or $B$:

(a) \emph{Node $i$ in network $A$}: one of its $k_i+1$ neighbors, node
$j$ with state $S_j$, is randomly chosen. If $i$ and $j$ share the
same orientation ($S_iS_j>0$), then with probability $p$ node $i$
adopts an extremist state if it is a moderate ($S_i=\pm 1 \to \pm 2$),
and, independently of the interaction, remains extremist if it is
already an extremist ($S_i=\pm 2 \to \pm 2$) [see
  Fig \ref{esquema}A]. If $i$ and $j$ have opposite orientations
($S_iS_j<0$), with probability $q$ node $i$ becomes moderate if it is
an extremist ($S_i=\pm 2 \to \pm 1$), or changes orientation if it is
a moderate ($S_i=\pm 1 \to \mp 1$) [see Fig \ref{esquema}B].

(b) \emph{Node $i$ in network $B$}: the state of $i$ changes with
probability
\begin{equation}
  P_B(S_i \mapsto -S_i)= \left(\frac{n}{k_i+1} \right)^{\beta},
  \label{PB}
\end{equation}
where $n$ is the number of neighbors of $i$ with opposite orientation
than $i$, and $\beta \ge 0$ is the volatility.

In the next Section we explore the behavior of the model using $\beta$
and $r$ as external control parameters. 
\\

\section{Simulation Results}
\label{simulation}

We studied the model described in Section~\ref{model} by means of
Monte Carlo simulations using two interconnected degree-regular random
networks (DR) of degree $\mu=5$ and $N$ nodes each.  We implemented
the Molloy-Reed algorithm \cite{Mol_01} to build the networks, where
each node is connected to $\mu$ random nodes in the same network, and
to one random node in the other network. Starting from a polarized
situation that consists of setting all nodes in network $A$ to
positive states and all nodes in network $B$ to negative states, we
let the system evolve following the M-model and the AS dynamics
described in Section~\ref{model} for networks $A$ and $B$,
respectively.  We investigated how the steady state of the system
depends on the opinion reinforcement $r$ and volatility $\beta$ that
control, respectively, the strength of agents' persuasion in network
$A$ and the likelihood that an agent in network $B$ changes its
decision.  Because we were particularly interested in studying whether
the dynamics in network $A$ prevails over the dynamics in network $B$
(or vice versa), we run many independent realizations of the dynamics
and calculated the probability $P_+$ that the entire two-network
system reaches a $+$ consensus, that is, the initial orientation
adopted by network $A$.  We consider that the system reaches consensus
when all nodes of both networks have the same orientation (either
positive $+$ or negative $-$). Notice that, for instance, states $S=2$
and $S=1$ are both considered as positively oriented.  The probability
$P_+$ was estimated as the fraction of realizations that ended in a
$+$ consensus.  Given that each separate model always reaches
consensus in a finite network --as explained in Section \ref{model}--,
one can check that the probability of a $-$ consensus in the entire
system is $P_-=1-P_+$.

\begin{figure}[H]

  \centering
  \includegraphics[width=\textwidth]{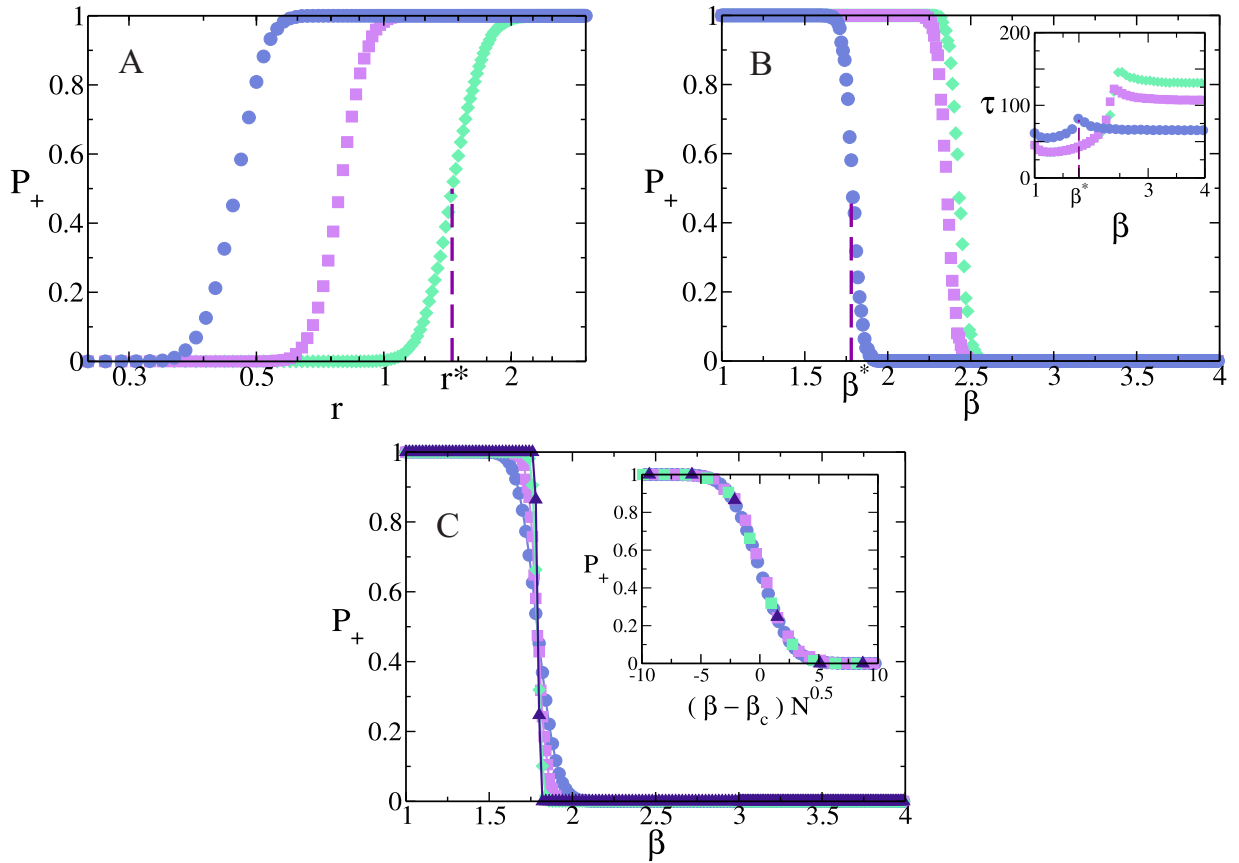}
  \vspace{0.4cm}
  \caption{\footnotesize Probability of positive consensus $P_+$ in a system of
    two interconnected networks $A$ and $B$.  Initially, all nodes in
    network $A$ ($B$) are positive (negative).  (a) $P_+$ as function
    of $r=p/q$ on a log-linear scale, for networks of size $N=2048$
    nodes and $\beta=2.0$ ($\circ$), $2.25$ ($\Box$) and $2.5$
    ($\Diamond$).  At the crossover point $r^*(\beta)$ is $P_+=1/2$
    (vertical dashed line shown for $\beta=2.5$ only).  (b) $P_+$ vs
    $\beta$ for $r=0.25$ ($\circ$), $1.0$ ($\Box$) and $1.2$
    ($\Diamond$).  At $\beta^*(r)$ is $P_+=1/2$ (vertical dashed line
    for $r=0.25$).  Inset: mean consensus time $\tau$ vs $\beta$, for
    the same parameter values, showing a maximum at $\beta^*$.  (c )
    $P_+$ vs $\beta$ for $r=0.25$ and network sizes $N=512$ ($\circ$),
    $2048$ ($\Box$), $8192$ ($\Diamond$) and $32768$
    ($\bigtriangleup$). Inset: the curves collapse when the $x$-axis
    is rescaled by $(\beta-\beta^*)\sqrt{N}$. All numerical results
    correspond to an average over $10^4$ independent realizations on
    degree-regular random networks of degree $\mu=5$.}
  \label{fig-P-r}
\end{figure}

In Fig \ref{fig-P-r}A we plot $P_+$ as a function of $r$ for three
different volatilities $\beta$. We observe that $P_+$ increases
abruptly from $0$ to $1.0$ when $r$ overcomes a crossover value
$r^*(\beta)$, determined as the symmetric point where $P_+=1/2$. This
means that for large reinforcement $r>r^*$ network $A$ imposes its
initial orientation to network $B$, and thus the dynamics of the
M-model prevails over the AS dynamics.  The opposite happens for low
reinforcement $r<r^*$, where the initial orientation of network B
prevails, and thus the dynamics of the AS model is stronger than that
of the M-model.  An interpretation of these results can be given in
terms of the response of the M-model to a variation in $r$.  As
described in Section \ref{model}, the initial positive consensus in
the M-model on network $A$ becomes more stable as $r$ increases.  Then,
it turns out that for very small values of $r$ the initial A-consensus
is very unstable, and all nodes in network $A$ quickly adopt the
negative states hold by nodes in network B, driving the entire system
to a $-$ consensus in most realizations ($P_+ \simeq 0$).  In the
opposite limit of very large values of $r$, the initial A-consensus is
very stable, thus most A-nodes keep their initial positive states
while B-nodes change their states to positive, and the entire system
reaches a $+$ consensus in most realizations ($P_+\simeq 1$).
Finally, for intermediate values of $r$ some realizations end in a $+$
consensus while the rest end in a $-$ consensus, leading to the
sigmoidal shape of $P_+$ vs $r$ in Fig \ref{fig-P-r}A.
\\

In Fig \ref{fig-P-r}B we plot $P_+$ vs $\beta$ for three values of
$r$.  We can see a crossover from $+$ to $-$ consensus at a value
$\beta^*(r)$, where $P_+=1/2$, in a similar fashion to the crossover
with $r$ described above.  For $\beta > \beta^*$ network $B$ imposes
its initial orientation to network $A$, while for $\beta < \beta^*$
the opposite happens.  This behavior can be explained using arguments
similar to those used above to explain the crossover of $P_+$ at
$r^*$. As $\beta$ increases from small values, the initial $-$
consensus state of network $B$ gains stability, continuously increasing
the probability that the system reaches a $-$ consensus or,
equivalently, decreasing $P_+$. The reason why curves start at
$\beta=1$ is because for low values of $\beta$ consensus states are
never observed in the simulations, even though finite systems must
reach consensus as we noted before. As we shall see when we analyze
other observable like the magnetization, for $\beta < \beta_c \simeq
0.86$ the system falls in an active steady state with $+$ and $-$
orientations coexisting in both networks but, after a long time,
consensus is eventually achieved by fluctuations. Consensus times in
this regime are extremely long for the system sizes we used, and thus
consensus is never achieved in a reasonable computer time.  Indeed, we
have run simulations on small enough networks and checked that an
absorbing state is always reached. As we shall explain, this
quasistationary non-consensus state is related to the coexistence
dynamics observed in the AS model for $\beta < \beta_c \simeq 1$.

Fig \ref{fig-P-r}C shows $P_+$ vs $\beta$ for $r=0.25$ and
different network sizes $N$.  We can see that the crossover becomes
sharper as $N$ increases, with a slope at $\beta^*$ that diverges as
$\sqrt{N}$, as the data collapse in the inset of Fig \ref{fig-P-r}C
shows. In the inset of Fig \ref{fig-P-r}B we show the mean time
$\tau$ to reach the consensus state as a function of $\beta$, for the
values of $r$ of the main Fig \ref{fig-P-r}B. We observe that
$\tau$ has a peak at $\beta^*$, which is consistent with the fact that
at the crossover point the system can reach either $+$ or $-$
consensus with the same probability $1/2$, suggesting that large
fluctuations lead the system to the final state. In Section~\ref{MF}
we give an insight into this last behavior and show that the breaking
in the symmetry of the system at $\beta^*$ eventually happens after a
long time, when finite-size fluctuations make the system overcome a
potential barrier. Below $\beta^*$ the M-model in network $A$ seems to
control the dynamics of the system --as there is a $+$ consensus in
both networks--, and thus $\tau$ is determined by the time it
  takes for network $B$ to reach a $+$ orientation from an initial $-$
orientation, which increases with $\beta$. But above $\beta^*$ the
opposite happens: network $B$ rules the dynamics, and thus $\tau$ is
related to the time that network $A$ takes to go from a positive to a
negative orientation. This observation is in agreement with the fact
that $\tau$ approaches a constant value as $\beta$ becomes large,
given that the M-model is independent of $\beta$, and then so is
$\tau$.

In order to explore the behavior of the system for a wider range of
$\beta$, we study the magnetization in networks $A$ and $B$, $m_A$ and
$m_B$, respectively, at the steady state. The magnetization in network
$\ell$ ($\ell=A,B$) at time $t$ is defined as
\\
\begin{equation}
  m_\ell=\sigma_{\ell}^+ - \sigma_{\ell}^-,
  \label{Eq.mi}
\end{equation}
with $m_{\ell} \equiv m_{\ell}(t)$, $\sigma_{\ell}^+\equiv
\sigma_{\ell}^+(t)$, $\sigma_{\ell}^-\equiv \sigma_{\ell}^-(t)$, and
where $\sigma_{\ell}^+$ and $\sigma_{\ell}^-$ are the fractions of
nodes with $+$ and $-$ state, respectively, in network $\ell$ at time
$t$.

\begin{figure}[H]
  \centering
  \includegraphics[width=0.7\textwidth]{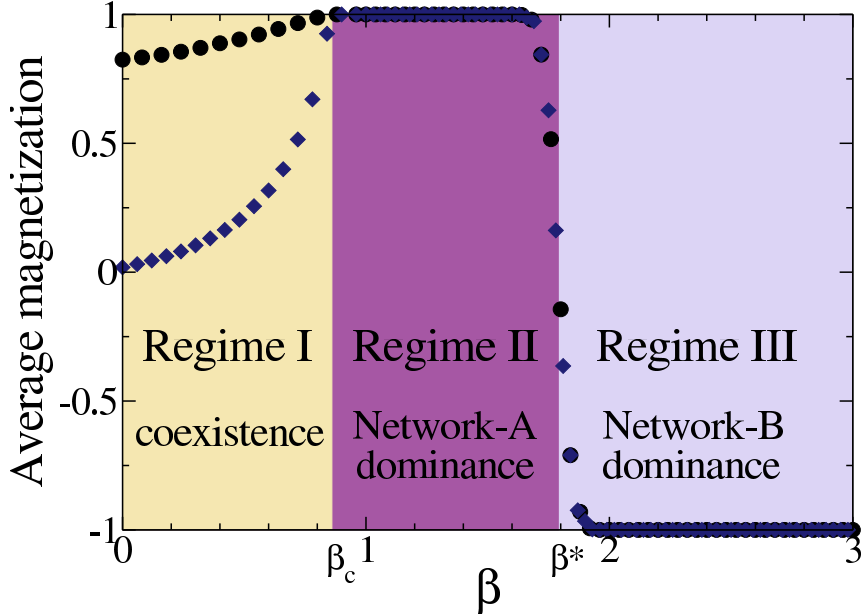}
  \caption{\footnotesize Average magnetization at the steady state $\langle m_A
    \rangle$ (circles) and $\langle m_B \rangle$ (diamonds) in networks
    $A$ and $B$, respectively, as a function of $\beta$, for $r=0.25$.
    Below the critical threshold $\beta_c \simeq 0.86$ the system
    remains in a disordered state where both $+$ and $-$ orientations
    coexist (Regime I), while above $\beta_c$ the system reaches an
    ordered state of consensus (Regimes II and III).  The point $\beta^*$ 
    denotes the crossover between Regimes II and III,
    characterized by a positive and negative consensus, respectively.
    Numerical results correspond to two DR random networks of degree
    $\mu=5$ and size $N=2048$ each, averaged over $10^4$ independent
    realizations.}
  \label{fig3}
\end{figure}

As we mentioned above, consensus in one of the two orientations is
only observed in the simulations when $\beta$ is above a critical
value $\beta_c \simeq 0.86$, while for $\beta<\beta_c$ the system
remains in an active steady state with both positive and negative
orientations coexisting.  This means that, in the $0 \le \beta <
\beta_c$ region, magnetizations $m_A$ and $m_B$ in a single
realization fluctuate around two different stationary values that are
neither $1.0$ nor $-1.0$.  This is shown in Fig \ref{fig3}, where we
plot the average magnetization over many realizations at the steady
state in each network, $\langle m_A \rangle$ and $\langle m_B
\rangle$, as a function of $\beta \ge 0$, for $r=0.25$ and
$N=2048$. We can distinguish three different regimes. In the first
regime (denoted by regime I), we see that $\langle m_A \rangle$
($\langle m_B \rangle$) increases from $0.8$ ($0.0$) to $1.0$ ($1.0$)
in the range from $\beta=0$ to $\beta_c \simeq 0.86$. That is, there
is a majority of nodes with positive orientation in network $A$, while
in network $B$ the coexistence is more even. We note that, strictly
speaking, this coexistence regime is stable only in the thermodynamic
limit, where the system remains forever in a stationary state of
non-consensus.  As stated before, in finite systems the steady state
lasts for very long times, but fluctuations ultimately drive the
system to an absorbing consensus state.

Above $\beta_c$ the system reaches a positive consensus $\langle m_A
\rangle=\langle m_B \rangle=1.0$ (network-$A$ dominance) for
$\beta_c<\beta<\beta^*$ (denoted by regime II), and a negative
consensus $\langle m_A \rangle=\langle m_B \rangle=-1.0$ (network-$B$
dominance) for $\beta>\beta^*$ (denoted by regime III). In regimes II
and III close to $\beta^*$, an average value of the magnetization
different from $1$ and $-1$ means that some fraction of the
realizations ended in a positive consensus and the rest in a negative
consensus.

The values of $\beta_c$ and $\beta^*$ are very different in
nature. While $\beta_c$ denotes a critical point from a disordered
phase (regime I) to an ordered phase (regimes II and III), $\beta^*$
denotes a crossover point within the ordered phase, which separates
the two dominance regions.  We also note that the order-disorder
transition at $\beta_c$ is related to the same type of transition
observed in the AS model, explained in Section \ref{model}. It seems
that the coexistence phase in the isolated AS dynamics is very robust,
and the coupling to the M-model produces only a shift in the critical
value, from $\beta_c \simeq 1$ to $\beta_c \simeq 0.86$.
\\

\begin{figure}[H]
  \centering
  \includegraphics[width=0.7\textwidth]{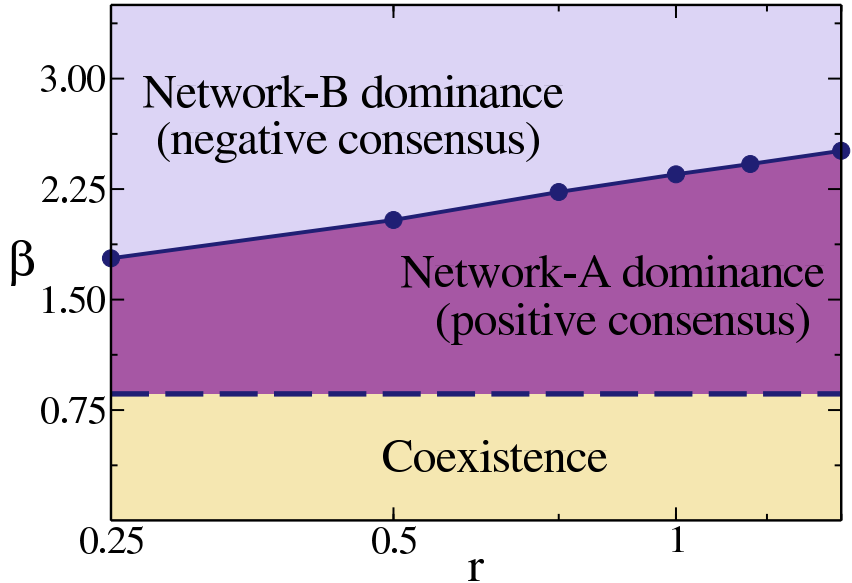}
  \caption{\footnotesize Reinforcement-volatility $(r-\beta)$ phase diagram on a
    log-linear scale for a two-network system with the same parameters
    as in Fig \ref{fig-P-r}. Solid circles correspond to the
    crossover points $(r^*,\beta^*)$ between network-A and network-B
    dominance regions, while the dashed line represents the transition
    point $\beta_c \simeq 0.86$ between coexistence and consensus.}
  \label{fig2}
\end{figure}

Fig \ref{fig2} shows the phase diagram of the system in the
$r-\beta$ plane, on a log-linear scale. We observe that the crossover
point $\beta^*$ increases very slowly (logarithmically) with $r^*$.
Therefore, starting from a point $(r,\beta)$ inside the B-dominance
region, an exponentially large increase in $r$ must be done to take
the system to the A-dominance region. In other words, for a small
change in the volatility of the decision making dynamics of network B,
the dynamics of network $A$ has to increase its opinion reinforcement by
a large amount, in order to impose its initial orientation.
\\

In the next Section we develop a theoretical approach that allows to
explain the qualitative behavior of the system in the three regimes.
Even though this approach assumes that the system is infinitely large,
is able to capture most of the phenomenology observed in the
simulations, which are for finite networks.

\section{Mean Field Approach}
\label{MF}

As we showed in Section~\ref{simulation}, the system exhibits three
different regions in the $r-\beta$ phase space: a coexistence of $+$
and $-$ nodes for $\beta$ below a critical value $\beta_c$, a $+$
consensus for $\beta_c < \beta < \beta^*(r )$ where the M-model in
network $A$ dominates, and a $-$ consensus for $\beta > \beta^*(r )$
where the AS model in network $B$ dominates.  In order to understand the
role of $\beta$ and $r$ in the behavior of the system in these three
regions, we study in this Section the evolution of the system within a
mean-field approach. To be specific, we write and analyze approximate
equations for the time evolution of the magnetization in each network.

As the system is symmetric at $\beta^*$, where consensus is equally
reached in both opinion orientations, we assume that the dynamics of
both models are equivalent at $\beta^*$ and, therefore, we consider
the M-model as an AS model with a volatility exponent
$\beta^*$. Roughly speaking, we can think of mapping the four-state
M-model into a two-state AS model by combining $S=1$ and $S=2$ states
into a single $+$ state and $S=-1$ and $S=-2$ into a single $-$ state,
and considering effective transition probabilities between $+$ and $-$
states that are non-linear functions of the fractions $\sigma^+$ and
$\sigma^-$ of $+$ and $-$ neighbors of a given node, respectively.
For instance, the effective transition probability of a node $i$ from
$-$ to $+$ can be written as $(\sigma^+)^{\beta^*}$, where $\sigma^+$
is the fraction of $i$'s neighbors in the opposite state $+$ ($S=1$
and $S=2$ states).  Even though it is difficult to obtain the exact
value of the exponent $\beta^*$, one can show that $\beta^*$ should be
larger than $1.0$ using the following heuristic argument.  The
effective transition probability from $-$ to $+$ states involves
single jumps from nodes in state $S=-1$ to state $S=1$, whose
probability is proportional to the fraction of $+$ neighbors
$\sigma^+$, and also double jumps from nodes in state $S=-2$ to $S=-1$
and then to $S=1$, with a probability proportional to $(\sigma^+)^2$.
Combining these two types of transitions in the entire network results
in an effective probability with an exponent $1.0 < \beta^* < 2.0$.

The advantage of mapping the four-state M-model into a two-state model
is that it allows to reduce the original two-network system --where the
M-model interacts with the AS model-- to a simpler system consisting
on two interacting AS models, which can be studied analytically.  Even
though these two systems are not exactly the same because the mapping
of the M-model into the AS model is only approximate, we shall see that
both systems share the same phenomenology, with results that are in
qualitative agreement with the simulation results of Section
\ref{simulation}, including a transition and a crossover between the
different regimes.

Based on these assumptions, we study a system that consists of two
interconnected networks $A$ and B, where an AS dynamics with fixed
volatility $\alpha=\beta^*$ runs on network $A$ (representing the
M-model), and another AS dynamics with variable volatility $\beta$
runs on network $B$. We start by deriving an approximate equation for
the time evolution of the magnetization
$m_\ell=\sigma_\ell^+-\sigma_\ell^-$ in network $\ell$ ($\ell=A,B$),
where $\sigma_\ell^S$ is the fraction of nodes with state $S$
($S=+,-$) in each network, which obeys the normalization condition
$\sigma_\ell^++\sigma_\ell^-=1$. At each time step $\Delta t=1/2N$, a
node $i$ in network $A$ with state $S$ is chosen with probability
$\sigma_A^S/2$, and switches to state $-S$ with probability $P_A(S \to
-S)$, changing $m_A$ by $\Delta m_A=-2S/N$.  Then, the average change
in the magnetization of network $A$ can be written as \\
\begin{equation}
  \frac{dm_A}{dt} = \frac{1}{1/2N}\left[\frac{\sigma_A^-}{2} \,
    P_A(-\to +) \frac{2}{N} - \frac{\sigma_A^+}{2}\,P_A(+ \to -)
    \frac{2}{N} \right]. 
  \label{dmAdt}
\end{equation}
\\ Using Eq.~(\ref{PB}) for the switching probability, $P_A$ can be
approximated as
\\
\begin{equation}
  P_A(S \to -S) \simeq \left( \frac{\left \langle n_A \right
    \rangle}{\mu+1} \right)^{\alpha}, 
  \label{PA}
\end{equation}
where $\left \langle n_A \right \rangle$ is the expected number
of neighbors of node $i$ with opposite state $-S$, and $\mu+1$ is the
total number of neighbors. Within a mean-field approach that neglects
nearest-neighbor correlations (node approximation), a neighbor of $i$
in network $A$ ($B$) is in state $-S$ with probability $\sigma_A^{-S}$
($\sigma_B^{-S}$) and, therefore, the expected number of
neighbors with state $-S$ of $i$ can be estimated as
\\
\begin{equation}
  \left \langle n_A \right\rangle \simeq \mu \, \sigma_A^{-S} +
  \sigma_B^{-S}. 
  \label{nA}
\end{equation}
 
Using Eqs.~(\ref{PA}) and (\ref{nA}) and expressing the densities
of states in terms of the magnetization $\sigma_A^S=(1+S\, m_A)/2$,
Eq.~(\ref{dmAdt}) can be written as
\\
\begin{equation}
  \label{mava}
  \frac{dm_A}{dt} = \frac{(1-m_A)}{2^{\alpha}(\mu+1)^{\alpha}} 
  \left[ \mu (1+m_A)+1+m_B \right]^{\alpha} -
  \frac{(1+m_A)}{2^{\alpha}(\mu+1)^{\alpha}} \left[ \mu (1-m_A)+
    1-m_B \right]^{\alpha},
\end{equation}

and a corresponding equation can be derived for $m_B$,
\\
\begin{equation}
  \label{mbvb}
  \frac{dm_B}{dt} = \frac{(1-m_B)}{2^{\beta}(\mu+1)^{\beta}} 
  \left[ \mu (1+m_B)+1+m_A \right]^{\beta} -
  \frac{(1+m_B)}{2^{\beta}(\mu+1)^{\beta}} \left[ \mu (1-m_B)+1-m_A \right]^{\beta}.
\end{equation}

Equations~(\ref{mava}) and (\ref{mbvb}) can be rewritten in the
form of a time-dependent Ginzburg-Landau equation \cite{Vazquez_10}
\\
\begin{eqnarray}
  \label{dmAdt2}
  \frac{dm_A}{dt} &=& - \frac{\partial V_A}{\partial m_A}\ , \\
  \label{dmBdt}
  \frac{dm_B}{dt} &=& - \frac{\partial V_B}{\partial m_B}\ ,
\end{eqnarray}
\\ with potentials $ V_A \equiv V_A(m_A,m_B)$ and $V_B \equiv
V_B(m_A,m_B)$ given by \\
\begin{eqnarray}
  V_A = &-& \frac{ \left[(\mu(1+m_A)+1+m_B \right]^{\alpha+1} 
    \Big\{ \mu \left[2+(\alpha+1)(1-m_A)\right] +1+m_B] \Big\} }
    {2^{\alpha}(\mu+1)^{\alpha} \mu^2\;(\alpha+1)\;(\alpha+2)} \nonumber \\
    &-& \frac{ \left[(\mu(1-m_A)+1-m_B \right]^{\alpha+1} 
      \Big\{ \mu \left[2+(\alpha+1)(1+m_A)\right] +1-m_B] \Big\} }
      {2^{\alpha}(\mu +1)^{\alpha} \mu^2\;(\alpha+1)\;(\alpha+2)}, ~~~~~~
      \label{VA}
\end{eqnarray}

\begin{eqnarray}
  V_B = &-& \frac{ \left[(\mu(1+m_B)+1+m_A \right]^{\beta+1} 
    \Big\{ \mu \left[2+(\beta+1)(1-m_B)\right] +1+m_A] \Big\} }
    {2^{\beta}(\mu+1)^{\beta} \mu^2\;(\beta+1)\;(\beta+2)} \nonumber \\
    &-& \frac{ \left[(\mu(1-m_B)+1-m_A \right]^{\beta+1} 
      \Big\{ \mu \left[2+(\beta+1)(1+m_B)\right] +1-m_A] \Big\} }
      {2^{\beta}(\mu +1)^{\beta} \mu^2\;(\beta+1)\;(\beta+2)}. ~~~~~~
      \label{VB}
\end{eqnarray}

This formalism is very useful for visualizing the system's evolution,
as each magnetization evolves towards the minimum of its associated
potential. However, unlike it happens in the AS model on a single
isolated network \cite{Vazquez_10} where the potential depends on a
unique magnetization and is static, the present case has two coupled
potentials that vary in time. Indeed, Eq.~(\ref{VA}) for the
potential $V_A$ that rules the evolution of $m_A$ can be interpreted
as an explicit function of $m_A$, whose shape is controlled by a
time-dependent external parameter $m_B$. Therefore, the shape of $V_A$
varies with time through $m_B$. An analogous interpretation can be
done for $V_B$, which depends on $m_A$. Thus, within this approximate
mathematical formalism represented by the coupled system of
Eqs.~(\ref{dmAdt2}) and (\ref{dmBdt}), the interplay between both
networks enters through the potentials $V_A$ and $V_B$, which interact
and co-evolve in time.

We now explore the behavior of the two networks by studying the
evolution of the magnetization described by Eqs.~(\ref{mava}) and
(\ref{mbvb}), and using the potential formalism. For network $A$ we set
the volatility value $\alpha=\beta^*=1.78$ corresponding to the
crossover point for $r=0.25$ calculated in Section~\ref{simulation},
and vary the volatility $\beta$ in network $B$. The phenomenology
described below is qualitatively the same for the values of $\alpha$
that correspond to the other values of $r$ used in Fig \ref{fig2}.

\begin{figure}[H]
  \centering
  \includegraphics[width=\textwidth]{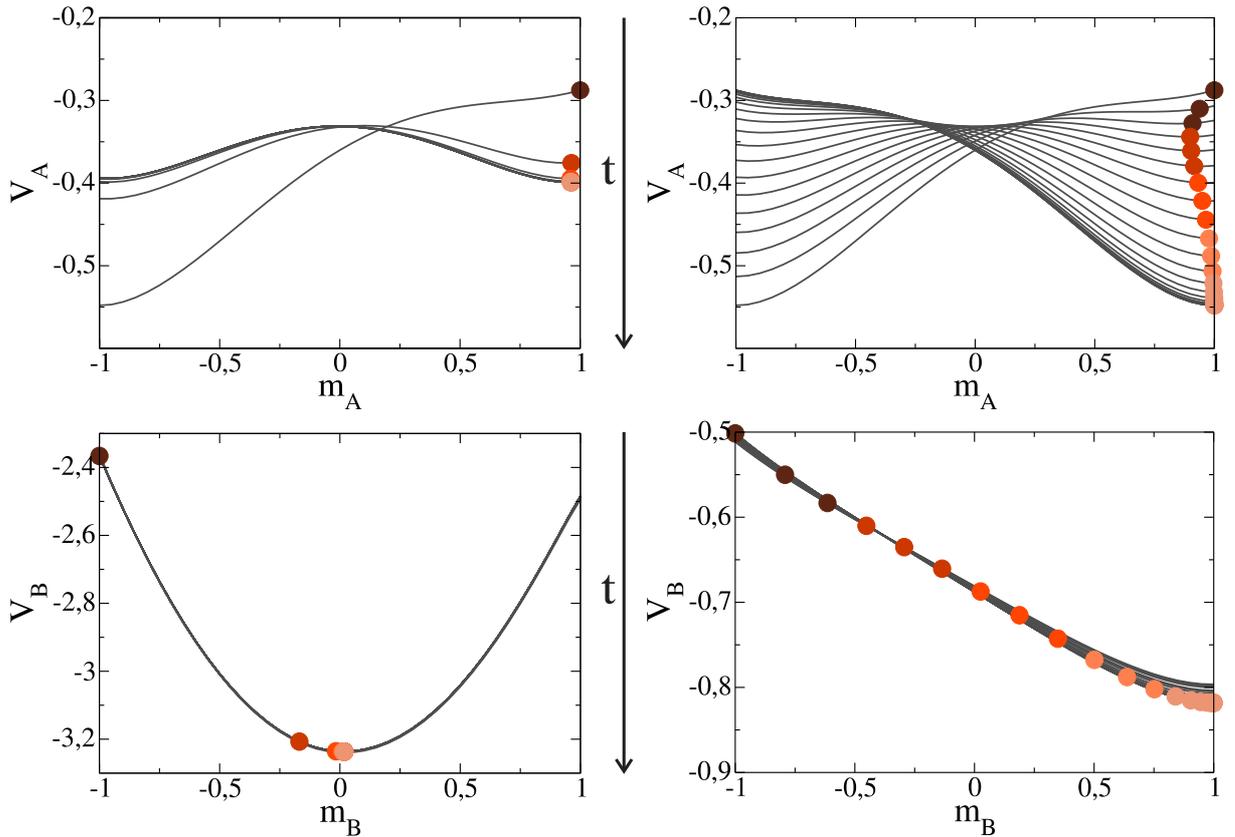}
  \caption{\footnotesize Potentials $V_A$ and $V_B$ (solid lines) as a function of
    the magnetizations $m_A$ and $m_B$ in networks $A$ and B,
    respectively, at different times, obtained from Eqs.~(\ref{VA})
    and (\ref{VB}).  The degree in both networks is $\mu=5$.  The
    volatility in network $A$ is $\alpha = 1.78$, while in network $B$
    is $\beta=0.1$ (left panel) and $\beta=1.2$ (right panel). 
    Circles correspond to the values of the magnetizations at
    different times, starting from the dark topmost circle at $t=0$
    and ending at the lightest circle for long times.  Vertical arrows
    indicate the time direction. Plots in the left panel show the
    coexistence regime I, while plots in the right panel describe the
    positive consensus regime II.}
  \label{fig4}
\end{figure}

\begin{figure}[H]
  \centering
  \includegraphics[width=\textwidth]{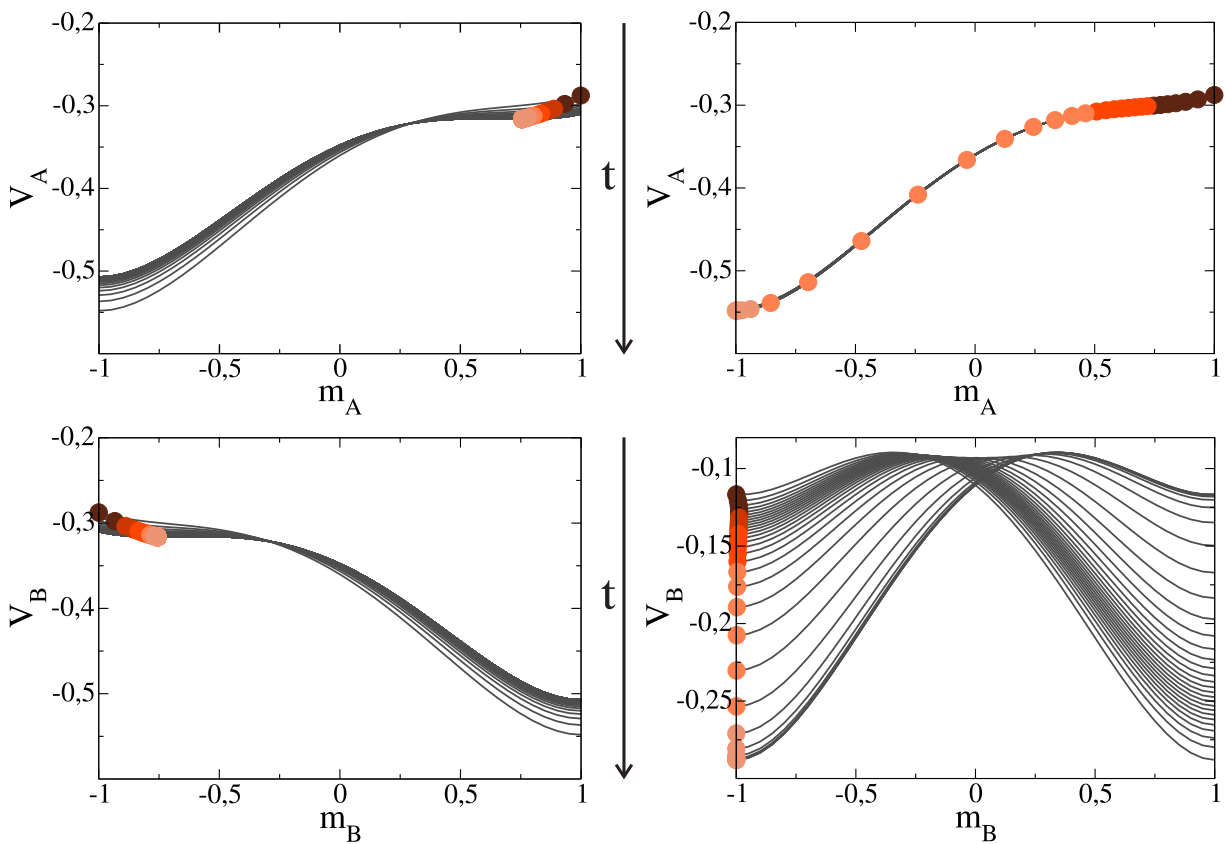}
  \caption{\footnotesize Potentials $V_A$ and $V_B$ as in Fig \ref{fig4}, but for
    volatility values $\beta=\alpha=1.78$ (left panel) and $\beta=3$
    (right panel). Plots in the left panel correspond to the crossover point, a
    symmetric case where the system remains disordered, while plots in the right
    panel show the negative consensus regime III.}
  \label{fig5}
\end{figure}

To visualize the trajectories of the magnetization, we plot in
Figs \ref{fig4} and \ref{fig5} the values of $m_A$ and $m_B$
(circles) and their associated potentials (solid lines) at different
times, for various parameter values. Each circle corresponds to the
magnetization $m_{\ell}$ at a given time $t$, which lies over the
potential $V_{\ell}$ at the same time $t$, with $\ell=A, B$. The
intensity of a circle's color decreases as time increases, starting
from $t=0$ (dark circle) and ending at the lightest color. Drawing the
complete shape of the potential helps to understand the trajectory
followed by $m_{\ell}$, which moves in the direction of the minimum of
$V_{\ell}$. The values of $m_A$ and $m_B$ were obtained by integrating
numerically Eqs.~(\ref{mava}) and (\ref{mbvb}), while the potential
$V_{A}$ at a given time $t$ was drawn by replacing the value of $m_B$
into Eq.~(\ref{VA}), and similarly for $V_{B}$.

Fig \ref{fig4} (left) shows the behavior in the coexistence regime I,
for $\beta=0.1 < \beta_c \simeq 0.86$. As we can see, the
magnetization in network $B$ evolves from $m_B=-1.0$ at $t=0$ to the
minimum at $m_B \simeq 0$ for long times (approximately $51 \%$ of
positive agents), while $m_A$ in network $A$ starts at $1.0$ and
reaches the stationary value $m_A \simeq 0.95$ close a positive
consensus. This result is in agreement with the one found from
simulations for small $\beta \leq \beta_c$ (see Fig \ref{fig3} for
small $\beta$), where the system remains in a disordered phase with a
coexistence of both orientations.   The behavior in the positive
consensus regime II is quite different [Fig \ref{fig4} (right)].
There we use $\beta=1.2$ that lies between $\beta_c \simeq 0.86$ and
$\beta^*=1.78$.  We observe that, as it happens in simulations,  both
networks reach a positive consensus after a few time steps.  While
$m_A$ quickly gets trapped in a local minimum that ultimately reaches
the value $m_A=1$, $m_B$ follows a direct trajectory from $m_B=-1$
towards a unique minimum at $m_B=1$.   The critical value of $\beta$
that separates regime I (coexistence) from regime II (consensus) was
found to be close to $1.0$ (not shown), which is quite different from
the critical threshold $\beta_c \simeq 0.86$ obtained from Monte Carlo
simulations.  This discrepancy may be due to the fact that the
theoretical approach considers an AS model in network $A$ (instead of
the M-model) and also that Eqs.~(\ref{mava}) and (\ref{mbvb}) describe
the evolution of $m_A$ and $m_B$ in infinite large systems, as they do
not have any terms that take into account finite-size fluctuations.
\\

Fig \ref{fig5} (left) corresponds to the crossover point
$\beta=\beta^*=\alpha$. We see that the magnetizations reach the
stationary values $m_A \simeq 0.75$ and $m_B \simeq -0.75$,
corresponding to a totally symmetric case in which there is an
unbalanced coexistence of orientations in each network. Even though the
total magnetization $m_A+m_B=0$ at the crossover point agrees with the average
magnetization obtained from simulations (see Fig \ref{fig3}), there
is a discrepancy with simulations results, where consensus in one of
the two orientations is always obtained for each individual
realization due to finite-size fluctuations. This is because
Eqs.~(\ref{mava}) and (\ref{mbvb}) describe an infinite large system
where fluctuations are neglected and, therefore, the system can never
escape from the minimum. Due to the symmetry in both potentials, one
would expect a $50\%$ chance to escape towards either 
consensus state if fluctuations were present, which is consistent with
the equal consensus probability in each state $P_+=P_-=1/2$ shown in
section \ref{simulation}.  Finally,
Fig \ref{fig5} (right) corresponds to regime III, with
$\beta=3>\beta^*$. The behavior in this case is analogous to the one
of Fig \ref{fig4} (right), but with an ultimate negative consensus in
both networks ($m_A=m_B=-1$), in agreement with simulation results of
Section~\ref{simulation}.
\\

In summary, the theoretical approach of this Section allows to
understand the underlying behavior of the system in the different
regimes, and gives an insight into why a dynamics prevails over the
other.

\section{Discussion}
\label{conclusion}

In this work, we explored the interplay between two different
dynamical processes that take place on two interconnected networks $A$
and $B$.  The dynamics on network $A$ corresponds to the one of the
M-model for opinion formation with four states ($M=2$), which
implements the mechanisms of compromise and persuasion related by a
reinforcement parameter $r$. In network $B$ the dynamics is akin to
that of the Abrams-Strogatz model for decision making, with two states
and a volatility parameter $\beta$.  Both models have positive and
negative opinion orientations.  We initially set the system in a
symmetric condition, where all nodes in network $A$ have positive
states and all nodes in network $B$ have negative states, and studied
the conditions under which one of the two dynamics dominates. We found
that for a reinforcement larger than a crossover value $r^*(\beta)$
the dynamics on network $A$ dominates, as a positive consensus is
reached in both networks, while the opposite outcome is obtained for
$r < r^*(\beta)$ (network $B$ dominates).  As we have shown, this is
due to the fact 
that increasing the level of opinion reinforcement in network $A$
beyond a value $r^*$ produces a large number of positive extremists
that are able to resist the change of orientation, imposing their
positive orientation to the entire system.  Besides, the study of the
full $r-\beta$ phase space revealed a transition at a critical
threshold $\beta_c$, from a disordered phase where both orientations
coexist to an ordered phase characterized by a consensus of one of the
two orientations.   We also showed that both dynamics are equivalent
along the crossover line $(r^*,\beta^*)$ that separates the
A-dominance and B-dominance regions, as the consensus probability in
either state is the same on the $(r^*,\beta^*)$ line.  Taking
advantage of this symmetry, we developed a mean-field approach for the
evolution of the magnetization in each network, using a time-dependent
Ginzburg-Landau equation. This approach was able to reproduce
qualitatively the different regimes observed in the simulations, and
gave an insight into when and how the dominance of one dynamics takes
place.

In practical terms, the equivalence between both dynamics means that a
rather complex M-model with four opinion states and a reinforcement
$r^*$ can be mapped to a simpler two-state model with effective
transition probabilities given by the exponent $\beta^*(r^*)$.  This
mapping might be very useful to gain an analytical insight into the
behavior of the M-model, given that the dynamics of the two-state
equivalent model can be understood in terms of its associated
Ginzburg-Landau potential.  Despite the fact that this result is
particular of the opinion and decision making models used in this
work, we expect that analogous behaviors can be obtained using other
types of dynamics, beyond socially inspired models.   As a general
remark, one can argue that it is possible to gain a better
understanding of a complex and poorly known dynamics by coupling this
dynamics to a much simpler a better known two-state model, using two
similar interconnected networks as the underlying topology.

While our results are obtained using degree-homogeneous networks, it
might be worthwhile to study the system using different network
topologies, as real social networks are known to be quite
heterogeneous.  Even though we limited our internetwork topology to a
single random interlink per node, the addition of targeted interlinks
connecting specific nodes in both networks may bring new
phenomenology.  It could also be interesting to investigate how the
number of different opinion states in the M-model affects the results,
given that a more robust polarized state is expected as the maximum
opinion value $M$ increases.

\end{document}